\documentclass{article}

\usepackage{titling}
\title{IoTNetSim: A Modelling and Simulation Platform for End-to-End IoT Services and Networking}

\usepackage[sc]{mathpazo} 
\usepackage[T1]{fontenc} 
\usepackage{microtype} 
\usepackage[hmarginratio=1:1,top=25mm,columnsep=20pt]{geometry} 
\usepackage[hang, small,labelfont=bf,up,textfont=it,up]{caption} 
\usepackage{booktabs} 
\usepackage{float} 
\usepackage{paralist} 

\usepackage{color}
\usepackage{colortbl}
\usepackage[usenames,dvipsnames,table]{xcolor}
\usepackage{siunitx}
\usepackage{pifont}
\newcommand{\cmark}{{\color{blue!80}\ding{51}}}
\newcommand{\xmark}{{\color{black!60}\ding{53}}}
\definecolor{Gray}{gray}{0.92}

\usepackage{abstract} 

\usepackage{titlesec} 
\renewcommand\thesubsection{\Roman{subsection}} 
\titleformat{\section}[block]{\large\scshape\centering}{\thesection.}{1em}{} 
\titleformat{\subsection}[block]{\large}{\thesubsection.}{1em}{} 

\usepackage{fancyhdr} 
\usepackage{amsfonts,amsmath,amssymb,amstext,latexsym}
\usepackage{multirow}

\usepackage{graphicx}
\usepackage[caption=false]{subfig}
\newcommand{\fig}[1]{Fig.~\ref{#1}}

\usepackage[pdftex,
            pdfauthor={Yehia Elkhatib},
            pdftitle={IoTNetSim: A Modelling and Simulation Platform for End-to-End IoT Services and Networking},
            pdfkeywords={IoT, Internet of Things, Cloud computing, Fog computing, Micro-clouds, Mini-clouds, Cloudlets, simulation}]{hyperref}

\newcommand*{\sect}[1]{\S\ref{#1}}
\usepackage{xspace}

\newcommand*{\eg}{\textit{e.g.}\@\xspace}
\newcommand*{\ie}{\textit{i.e.}\@\xspace}
\makeatletter
\newcommand*{\etc}{%
    \@ifnextchar{.}%
        {etc}%
        {etc.\@\xspace}%
}
\newcommand*{\etal}{%
    \@ifnextchar{.}%
        {et al}%
        {et al.\@\xspace}%
}
\makeatother

\begin{document}

\date{}
\title{\thetitle}

\author{
	Maria Salama,
	Yehia Elkhatib,
	Gordon S. Blair\\
    School of Computing and Communications, Lancaster University, United Kingdom\\
    \normalsize Email: \href{mailto:y.elkhatib@lancaster.ac.uk}{y.elkhatib@lancaster.ac.uk}\\[4mm]
    \textbf{\textcolor{red}{This is a pre-print}}\\
    \textbf{\textcolor{red}{The final version is available on ACM DL}}\\
}

\maketitle

\thispagestyle{fancy} 

\begin{abstract}
Internet-of-Things (IoT) systems are becoming increasingly complex, heterogeneous and pervasive, integrating a variety of physical devices and virtual services that are spread across architecture layers (cloud, fog, edge) using different connection types. As such, research and design of such systems have proven to be challenging. Despite the influx in IoT research and the significant benefits of simulation-based approaches in supporting research, there is a general lack of appropriate modelling and simulation platforms to create a detailed representation of end-to-end IoT services, \ie from the underlying IoT nodes to the application layer in the cloud along with the underlying networking infrastructure. To aid researchers and practitioners in overcoming these challenges, we propose IoTNetSim, a novel self-contained extendable platform for modelling and simulation of end-to-end IoT services. The platform supports modelling heterogeneous IoT nodes (sensors, actuators, gateways, \etc) with their fine-grained details (mobility, energy profile, \etc), as well as different models of application logic and network connectivity. The proposed work is distinct from the current literature, being an all-in-one tool for end-to-end IoT services with a multi-layered architecture that allows modelling IoT systems with different structures. We experimentally validate and evaluate our IoTNetSim implementation using two very large-scale real-world cases from the natural environment and disaster monitoring IoT domains.
\end{abstract}

\section{Introduction}
\label{sec_introduction}
Internet-of-Things (IoT) systems and networks are increasingly becoming large, complex, heterogeneous and pervasive. They integrate a large variety of physical devices (IoT devices and sensors) communicating through different networking connections (cellular, WiFi) spread across different architecture layers (cloud, fog, edge). That is, IoT systems are spanning both virtual and physical domains.  The research process in IoT, starting with the idea formulation and culminating with real-world deployment, requires developing and validating initial proofs-of-concept and subsequent prototypes \cite{Chernyshev2018}. 

Given the large scale and heterogeneity of IoT systems and networks, designing and testing IoT services are challenging tasks \cite{Buyya2009} \cite{Armbrust2010}. Prototyping using a large number of hardware nodes may not be practical during the initial design phase. Similarly, benchmarking and setting up reproducible experiments are challenging undertaking tasks \cite{Calheiros2011} \cite{Chernyshev2018}. To this extent, we argue that simulation-based approaches are significantly important for research benchmarking, designing, testing and experimenting IoT systems and networks (\sect{sec_background}). 

Simulation-based approaches offer significant benefits to researchers and practitioners \cite{Quiroz2009} \cite{Calheiros2011}, supporting and accelerating research and development of systems, applications and services \cite{Buyya2009}. Simulation tools are generally important and necessary tools designed and developed to aid researchers in testing their hypothesis, benchmarking studies in a controlled environment and easily reproducing results, conducting experiments with different workloads and resource provisioning scenarios, as well as testing systems performance \cite{Quiroz2009} \cite{Buyya2009}. In the context of cloud computing, simulators have accelerated its research and development \cite{Buyya2009}, as quantifying the performance of service provision in real cloud environments is challenging \cite{Calheiros2011}. In IoT systems and networks, simulation tools have also claimed their importance to fill the gap between conceptual research and proof-of-concept implementation \cite{Chernyshev2018}. 

Despite the influx of research in IoT and the various simulations environments proposed so far, there is a general lack \textemdash to the best of our knowledge \textemdash of modelling and simulation environments to create a detailed representation of end-to-end IoT services and related networking. The recent survey by Chernyshev \etal \cite{Chernyshev2018} concluded that generally there is no available all-in-one simulator for end-to-end IoT services (\sect{sec_relatedWork}).

In this paper, we propose a novel self-contained platform for modelling and simulating end-to-end IoT services with detailed representation of IoT systems and networking components, namely \textit{IoTNetSim}. The main research objective is to assist researchers and practitioners in designing, validating and experimenting IoT systems and networks. To this extent, the proposed platform is designed as a multi-layered architecture, which allows modelling and simulating IoT systems with different structures, application models, IoT services and network connections (\sect{sec_architecture}). The modularity of the architecture allows modelling model systems with any combination of cloud, fog, edge, IoT components according to the system architecture and design. The extendable design supports modelling and testing bespoke IoT nodes and network types, as well as placement algorithms used in designing IoT systems (\sect{sec_design}). 

IoTNetSim platform contributes to the research community with: (i) detailed modelling of IoT nodes and sensors, including power sources and mobility, (ii) modelling and testing IoT networking, covering different types of network connections used in IoT systems, and (iii) modelling and simulation of IoT services and applications from the sensing data phase to data analysis in the cloud. The platform also supports modelling domain-specific IoT applications and end-to-end services, as well as processing and testing the performance of IoT systems under varying dynamic workloads with different quality goals.

We validate and evaluate the proposed platform with a series of experiments using two contrasting IoT case studies derived from active real-world deployments (\sect{sec_evaluation}, \sect{sec_discussion}). Specifically, we simulated an IoT testbed for the natural environment deployed in UK and a large-scale IoT testbed deployed in Japan. Adopting ``Open Science" principles, we release IoTNetSim environment as open-source software (Appendix~\ref{app_onlineResources}).


\section{Background}
\label{sec_background}
IoT is defined as ``a paradigm that considers pervasive presence in the environment of various things that through wireless and wired connections are able to interact and cooperate with other connected things to create seamless communication and contextual services and reach common goals" \cite{Misra2017}. Accordingly, the IoT paradigm spans both virtual and physical domains and is supported by the cloud, fog and edge computing paradigms \cite{AlFuqaha2015}. An IoT system is typically composed of several physical devices, spread across different layers (cloud, fog, edge), that carry out various tasks such as sensing, controlling, visualisation, data analysis, \etc.

The highly heterogeneous entities are interconnected using different connection types and form different network topologies \cite{Misra2017}. Data is communicated through these networks for analysis and decision making in the cloud with minimal or no human intervention \cite{AlFuqaha2015} \cite{Chernyshev2018}. Such connectivity also allows IoT devices to coordinate more complex tasks and exhibit self-$X$ capabilities such as configuration, optimisation, management, and adaptation \cite{Vermesan2011} despite many being resource-constrained in terms of processing power, memory capacity, or battery life.

Research and development in IoT are challenging due to the inherent nature of IoT integrating the physical world with the virtual one. A typical cycle of IoT research process starts with the formulation of an idea, eventually culminating in a real-world deployment. This process requires to develop and validate the initial proofs-of-concept and subsequent prototypes. Given the large scale of IoT services, prototyping using a large number of hardware nodes may not be practical during the initial exploratory design and evaluation phase, due to economic and operational constraints, especially when the reliability and utility of the protocol under consideration are yet to be proven \cite{Papadopoulos2017}. Additionally, setting up reliable and reproducible experiments involving real hardware is challenging and costly, and often requires specific expertise and domain knowledge~\cite{Papadopoulos2013}. As such, proof-of-concept and theoretical analysis are typically realised using simulations, and subsequent prototypes are experimentally evaluated in testbeds.

Generally, IoT simulators should capture precisely real-world complexity. This includes modelling and simulating the high-degree of heterogeneity in IoT nodes, diverse application domains and capturing diversity across IoT layers \cite{Kecskemeti2017}. IoT simulators are also challenged to offer high fidelity for real scenarios comprising nodes failures, network disconnections. Such tools should also support large-scale simulations and extensions custom requirements such as new protocols evaluation, IoT services \cite{Chernyshev2018}. Given the crucial role of networks in IoT ecosystems, networking should be expressively modelled, covering different types of networks, signals and protocols \cite{Papadopoulos2017}.

\section{IoTNetSim Architecture}
\label{sec_architecture}
In this section, we outline the architecture of \emph{IoTNetSim}, including the different layers for simulating IoT services and networking. 

\begin{figure}[!th]
\centering
\includegraphics[width=0.9\textwidth, clip, trim=0 0.2cm 0 0.1cm]{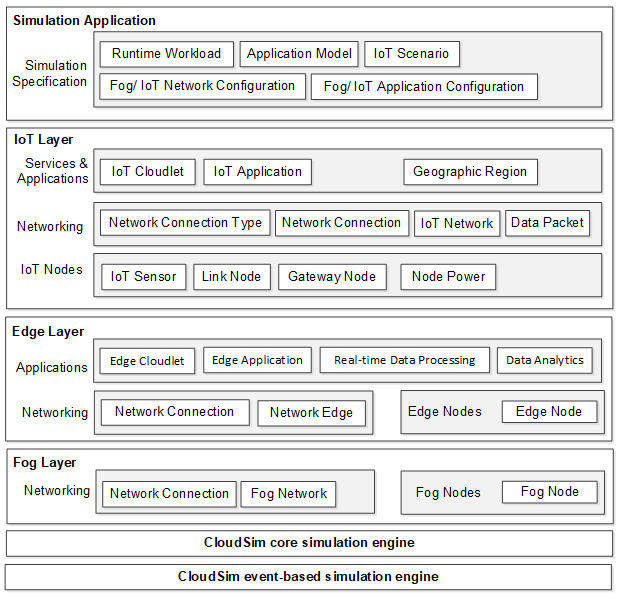}
\caption{The architecture of IoTNetSim.}
\label{fig_IoTsimArchitecture}
\end{figure}

The main objective of \emph{IoTNetSim} is modelling and simulating end-to-end IoT services with detailed representation of the IoT paradigm and associated connectivity. The design rationale of architecting \emph{IotNetSim} is a multi-layered modular architecture, which allows modelling and simulating different structures of IoT systems. For instance, a simple IoT system with a set of sensors connected to the cloud could be modelled with fine-grained details of the nodes. Meanwhile, a complex system with the sensors connected through edge and fog devices could also be modelled, allowing different topologies of networks. Following this design principle, the architecture of \emph{IoTNetSim} is mainly composed of different layers for the Cloud, Fog, Edge, IoT, Application Model and Simulation Application, built on top of an event-based simulation engine. \fig{fig_IoTsimArchitecture} shows the multi-layered architecture of \emph{IoTNetSim}. The multi-layered modular architecture supports modelling different structures for IoT systems and networks, as well as extensions for further functionalities and placement algorithms. Below, we discuss the details of each layer.

\paragraph*{\textbf{Event-based Simulation Engine}}
As shown in the figure, we have inherited the event-based simulation engine of \textit{CloudSim} \cite{Buyya2009} \cite{Calheiros2011}. The \textit{CloudSim} simulation toolkit is one of the widely-used general purpose cloud simulation environments \cite{Kathiravelu2014a} and the most sophisticated discrete event simulator for clouds \cite{Garg2011} (as discussed in \sect{sec_relatedWork}).

\paragraph*{\textbf{Cloud Layer}}
This layer extends the CloudSim core simulation engine. CloudSim defines the core entities of a cloud environment, such as datacenters, hosts physical machines (PMs), virtual machines (VMs), applications or user requests (called cloudlets) \cite{Buyya2009} \cite{Calheiros2011}. A \textit{Datacenter} is the resources provider simulating the infrastructure of the cloud (IaaS), which includes the hosts running virtual machines responsible for processing end-user requests (SaaS). Computational capacities of PMs and VMs (CPU unit) are defined by $Pe$ (Processing Element) in terms of million instructions per second (MIPS) \cite{Buyya2009} \cite{Calheiros2011}. Processing elements in a PM are shared among VMs and requests in a VM. The simulation also takes into account the memory, storage and energy consumption of different computational resources. A \textit{Datacenter Broker} is responsible for the allocation of end-user requests to VMs. Once the simulation is started, the requests are scheduled for execution, and the cloud behaviour is simulated. The cloud layer covers the typical components of a cloud, including the physical resources (datacenters and their PMs), virtual resources (VMs), virtual services (VM services for end-user VMs provisioning) and end-user services (for running their service requests and cloudlets).

\paragraph*{\textbf{Fog and Edge Layers}}
These layers model fog- and edge-related components respectively, where simulations could include either, both, or neither according to the conducted experiments. The fog layer includes details of fog-enabled nodes, fog networks configurations and different types of network connections. The edge layer, similar to the fog layer, includes edge-enabled nodes and network edges. This layer also includes edge cloudlets, edge applications, and modules for basic and real-time data processing.

\paragraph*{\textbf{IoT Layer}}
This layer encompasses different types of IoT nodes, networks and services. IoT nodes include fixed and mobile sensors, link nodes, and gateway nodes. The power sources of each node type are modelled in detail to allow the simulation of real-time scenarios. More details in \sect{sec_design_implementation}.

\paragraph*{\textbf{Simulation Application}}
This is the top-most layer for setting up different policies and simulation parameters, including scheduling policies, service types, application models, networking configurations, and runtime workloads and scenarios.

The multi-layered architecture allows implementing different functionalities of IoT nodes for processing and analysing data. For the networking in the different layers, details for data packets and network connection types are modelled for real-time scenarios of network loss and disconnection, as well as network traffic. The simulations results include quantitative measures of utility for simulating an end-to-end IoT service across all architectural layers. Researchers and practitioners, willing to design an IoT ecosystem and network or study the efficiency/ improvements of an existing one, would create instances of these layers with their design and IoT nodes to be used, along with network and applications configurations.

\section{Design and Implementation}
\label{sec_design}
In this section, we provide details related to the design classes and implementation of \textit{IoTNetSim}. 
\vspace{-0.5em}

\subsection{IoTNetSim Design}
\label{sec_design_classes}
\fig{fig_IoTSimClassDiagram} shows the class diagram of our design and implementation. The top-level design is composed of packages, which correspond to the layered architecture (\ie IoT, Edge, Fog, Cloud). The packages also allow encapsulation and reusability of entities that are common among different layers (\eg networking, nodes). In more details, the design is composed of the following packages:

\begin{figure*}[!t]
\centering
\includegraphics[width=\textwidth, clip, trim=0 0.375cm 0 0.3cm]{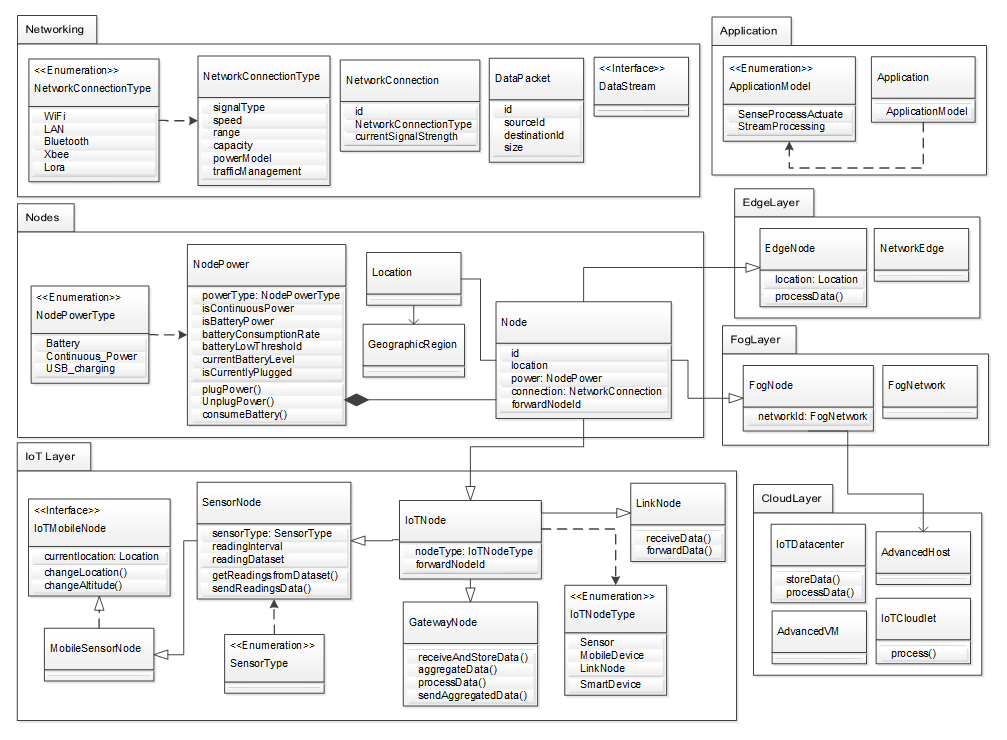}
\caption{The structure of IoTNetSim as a class diagram.}
\label{fig_IoTSimClassDiagram}
\end{figure*}

\textbf{Networking} is a common package among the different layers of the architecture. Each connection type is modelled with signal type, range, capacity and power model. An instance of a network connection would inherit these properties from the type, with the strength of the current signal. This allows simulating real and hypothetical scenarios of network loss and disconnections. Data Packet and Data Stream simulate the data transferred with different sizes between different nodes.

\textbf{Node} is shared among the IoT, Edge, Fog and Cloud layers. The package includes fine-grained properties that could be used for modelling and simulating IoT, edge and fog nodes. Properties include the power source, location, as well as geographic coverage. Configuring the location allows simulating different placement possibilities of IoT, edge and fog nodes, as well as their networks, enabling among other things the investigation of node placement algorithms.

\textbf{IoT} encapsulates the components necessary for modelling and simulating an IoT ecosystem. It is composed of: (i)~a general class for IoT nodes, (ii)~an interface for mobile nodes, (iii)~a class for sensor nodes with communication functionalities, (iv)~link node class with functions for receiving and forwarding data, and (v)~a gateway node class for nodes capable to aggregate, process and send data. Our implementation includes the basic functionalities of these components.

\textbf{Edge} supports creating instances of edge devices with configurable computational and storage capacities, as well as configurable functions of data processing and sending data to either cloud or IoT nodes. 

\textbf{Fog} creates instances of configurable fog nodes and associated networking. A fog node could be any computing/storage device or micro-server. These devices inherit their computational configuration from the Physical Host of \textit{CloudSim} and fine-grained properties from the Node package. 

\textbf{Cloud} (as the name implies) represents cloud infrastructure components. The IoT Datacenter is extended from the \textit{CloudSim} Datacenter to include functionalities of storing, processing and analysing data for IoT services. These functionalities could be either basic or real-time and could be further extended for complex data processing. The Hosts (physical machines) and VMs have also been extended for tracking energy consumption. IoT Cloudlets and Service Requests are tailored to support special types of IoT service requests, such as in-field enquiries and real-time alerts.

The described modular design allows performing simulations of different possible topologies. For instance, an IoT system could be designed with an infrastructure directly connected to the cloud. Other possibilities could be including either or both fog and edge layers. A node in any layer could be pre-configured with a forwarding node, or dynamically configured during runtime. 

\subsection{The Simulation Process}
\label{sec_design_simProcess}
\begin{figure}[!t]
\centering
\includegraphics[width=0.7\textwidth, clip, trim=0 0.2cm 0 0.2cm]{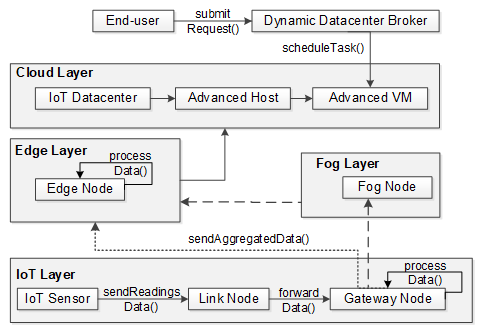}
\caption{The simulation process in IoTNetSim.}
\label{fig_simulationProcess}
\end{figure}

The platform is a discrete-event simulation with the flow of events illustrated in \fig{fig_simulationProcess}. The simulation process could start either by an end-user submitting a service request to the cloud where the broker will schedule and provide adequate resources or by IoT sensors regularly submitting their sensing data by the scheduled reading interval. Reading data are received by link nodes, which in turn forward it to the gateway node. A gateway node aggregates data received and sends it to either the fog, edge or cloud according to the configured IoT system architecture. The cloud resources are for data storage, processing and analysis. 

If edge nodes are present in the architecture, they will forward the data to the cloud and/or process it. In case of having fog nodes only, their job will be to forward data according to the configured topology. Nodes could also perform as \textit{actuators} to take actions according to the data received or the current state of the network.

\subsection{Implementation and Experiments Design}
\label{sec_design_implementation}

\paragraph*{\textbf{Modelling and Simulation of Cloud Infrastructure}}
We have extended some core classes of \textit{CloudSim}, namely IoTDatacenter, AdvancedHost and AdvancedVM. Extensions include adding necessary quality and power metrics. The IoTDatacenter is extended with functionalities of storing, processing and analysing data received from IoT nodes. The DatacenterBroker \textemdash responsible for workload distribution and resources provisioning \textemdash is also extended by queuing models necessary for runtime adaptation capabilities. 

\paragraph*{\textbf{Modelling and Simulation of IoT Nodes}}
Different types of IoT node could be modelled and configured, \ie sensors and connectivity nodes. 
A \textit{sensor node} is configured with reading interval and readings dataset. 
Connectivity nodes could be configured as link and gateway nodes. \textit{Link nodes} are responsible for receiving data from sensor nodes and forward it to \textit{gateway nodes}. The latter is more powerful nodes responsible for temporarily storing data received from link nodes. Gateway nodes can, then, aggregate data and send it to the cloud, and/ or process it for further actions. 
IoT nodes could also be configured as \textit{actuators} by implementing the required actions. Examples of actions include sending data to a new node, and adapting the flow of events or nodes configuration in case of node failure.  
Employing combinations of these various node types enables the modelling of all topologies and architectures of IoT systems.

IoT nodes are configured with a power source that could be a battery, USB charging point or continuous power supply. Battery consumption is tracked during the simulation for real deployments. Nodes are associated with different types of connections (\eg 3G WiFi) and tracked signal strength. Possible data of each sensor are stored in \texttt{csv} file, and the sensors are configured to read data from their files and submit a reading at each time interval. Data could be selected sequentially, randomly, or randomly within a specific range according to the hypothetical scenarios.

\paragraph*{\textbf{Support for IoT Nodes mobility}}
Examples of mobile sensors to be configured include animal tracking sensors and mobile phones. A \textit{mobile sensor} extends the sensor node with functionalities of changing location and altitude. Location and altitude are configured using $(x, y)$ and $z$ parameters respectively. These could be read from \texttt{csv} file, changed randomly or systematically following a specific pattern.

\paragraph*{\textbf{Modelling and Simulation IoT Services and Applications}}
We use the \textit{IoTServiceType} class to model IoT services offered by the IoT ecosystem, \eg monitoring service, alerting service or data acquiring service. A service type is configured by the computational resources it requires (MIPS). An \textit{IoTServiceRequest} and \textit{IoTCloudlet} are used to model an end-user request for a specific service type. This allows modelling dynamic workloads by multiple end-users for a variety of services. A \textit{RuntimeWorkload} is also added to allow conducting experiments for consecutive time intervals.

\paragraph*{\textbf{Support for Data Security and Privacy}}
Security and privacy are also considered for data transferred between nodes and the application layer. Our modular design allows researchers to implement their own algorithms.

\paragraph*{\textbf{Modelling and Simulation of Fog and Edge Computing}}
Physical devices could be enabled as fog or edge devices in an IoT system design. These devices are configured with a network connection and functions for data processing (cf.~\cite{elkhatib2017microclouds}). The platform allows designing the topology of a fog network and network edges where devices are located.

\paragraph*{\textbf{Modelling and Simulation of Networks and Connections}}
Different networks connections could be configured and associated with IoT, fog and edge nodes. Connection types include Wi-Fi, 3G, Bluetooth, LoRa, Zigbee, short- and long-range radio. Each connection is detailed with the signal type, capacity, power model and traffic management protocol. Parameters of any connection type could be obtained from models developed in the literature. For an IoT service, tracking the signal type for possible signal loss scenarios is important when evaluating systems design. Data sent by different nodes are encapsulated into data packets, where they are tracked across the network nodes along with their size.

\section{Experimental Validation and Evaluation}
\label{sec_evaluation}
We conducted experimental evaluation with the aims of: (i)~validating the components of the proposed platform; (ii)~evaluating its capability to simulate large-scale systems; and (iii)~assessing the IoTNetSim's capability to model and simulate different IoT systems. 
For the validation objective (aim i), we used a simple case study where we simulated an actual deployment of an end-to-end environmental IoT with different types of sensors and networking nodes (\sect{sec_evaluation_EnvIoT}). 
For the scalability evaluation (aim ii), we simulated the JOSE testbed for field trials of large-scale IoT services, which features a large-scale infrastructure and network for IoT services (\sect{sec_evaluation_JOSE}). 
Each case study features a different application model and IoT scenarios, and together they help to assess the generality of the platform by covering the whole design space (aim iii). 
We run all experiments on a PC with 2.7GHz Intel Core~i7 and 16GB RAM. 

\begin{table*}[!t]
\caption{Details of the nodes in the environmental IoT case study}
\label{tbl_EnvIoT_nodes}
\center
\footnotesize
\setlength{\tabcolsep}{0.5em}
\begin{tabular}
{
l
>{\raggedright}p{0.22\textwidth} 
>{\raggedright}p{0.24\textwidth} 
>{\raggedright\arraybackslash}p{0.34\textwidth} 
}
\toprule
{\textbf{Node}}             &     
{\textbf{Location}}         &     
{\textbf{Function}}         &     
{\textbf{Specification}} \\
\midrule
Fixed sensors                &
field of study                &
measure air temperature and rainfall        &
soil sensors with short-range radio 
\\
\rowcolor{Gray}
Mobile sensors                    &
attached to animals moving around the field of study                    &
measure location and attitude of animals            &
livestock tracker with GPS, accelerometer and short-range radio
\\

Relay node                &
field of study            &
receiving data from sensor nodes and sending them to gateway node            &
Raspberry Pi model A+ with short-range Xbee and long-range Xbee2 radio transmitter
\\
\rowcolor{Gray} 
Gateway node                &
$0.5km$ away from the field of study            &
store-and-forward data                        &
Raspberry Pi model B with a long-range radio transmitter and 3G cellular connection
\\
Cloud server            &
different place than the field of study            &
data storage                &
server with database and internet connection        
\\
\bottomrule
\end{tabular}
\end{table*}

\subsection{Case Study: End-to-end IoT for the Natural Environment}
\label{sec_evaluation_EnvIoT}
In a field study report \cite{Nundloll2016}, an end-to-end IoT infrastructure was designed and deployed for monitoring and managing the natural environment. The field study covers a geographic region around Conwy in North Wales, UK. This region is typical of many rural areas supporting industries including agriculture, forestry, tourism and fishing but facing challenges caused by climate change. The region has been exposed recently to many intensive flooding and pollution events, storms and periods of drought. An example scenario of the problem is the rejection of shellfish products by the European Union based on high microbiological pollutant load and risk to public health, where environmental IoT infrastructure has been used to identify anomalous events in the region. 

\begin{figure}[!h]
\centering
\includegraphics[width=0.98\columnwidth]{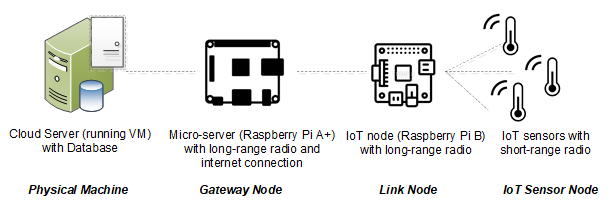}
\caption{An architectural view of the environmental IoT testbed (redrawn from \cite{Nundloll2016}).}
\label{fig_EnvIoTTestbed}
\end{figure}

\paragraph*{\textbf{Simulation Setup.}}
We have setup the deployment of this case study using our simulation environment. The testbed consists of: (i) sensors\footnote{We excluded animal mobile sensors in our initial evaluation due to the complexities of their movement patterns.} used for different environmental measurements, (ii) link nodes used to collect data received from sensor nodes and deliver them to the cloud in two kinds: long-range relay nodes and gateway nodes, and (iii) a cloud server with a database in which all data received from the gateway nodes are stored. Details of the nodes are listed in Table~\ref{tbl_EnvIoT_nodes}. Other cloud resources can then access this server to use the data for analysis, modelling or visualisation. For the day-to-day running of the deployment, the management service also records the last time it heard any data from each node, and the last-reported battery level of each node. The architecture of the testbed is illustrated in \fig{fig_EnvIoTTestbed}. 

\paragraph*{Environmental Data}
Since the actual readings were not available from the previous study \cite{Nundloll2016}, we configured the sensors to read from \texttt{csv} files of historical datasets from the UK Centre for Environmental Data Analysis \cite{MetOffice2017}.

\paragraph*{\textbf{Validation experiments and results.}}
The objective of the validation experiments is to examine the correctness of the events' flow of the application under investigation given the IoT architecture. In the validation experiments, we have setup the initial configuration of the testbed, as shown in Table \ref{tbl_EnvIoT_testbedConfiguration}. The testbed is composed of 3 air temperature sensors, 3 surface flow water sensors for air temperature and rainfall readings respectively, one relay node for forwarding data, one gateway node for aggregating, storing and forwarding data, as well as one cloud server for analysing data. Each sensor is configured to submit 4 readings/day (reading interval 6 hours) for 30 days. We have used readings of the most recent 30 days of the dataset (from December 2016). 

\begin{table}[!h]
\caption{Configuration of the environmental IoT testbed}
\label{tbl_EnvIoT_testbedConfiguration}
\center
\footnotesize
\begin{tabular}
{l r}
\toprule
{\textbf{Node Type}}             &     
{\textbf{Number of Nodes}}             
\\
\midrule
Air temperature sensor                &
3                
\\
\rowcolor{Gray} 
Surface flow water sensor        &
3
\\
Relay node                    &
1
\\
\rowcolor{Gray} 
Gateway node                &
1
\\
Physical host                &
1
\\
\midrule
\textit{Total}                            &
9
\\
\bottomrule
\end{tabular}
\end{table}

\begin{figure}[!h]
\centering
\includegraphics[width=0.8\textwidth, clip, trim=0 0.45cm 0 0.1cm]{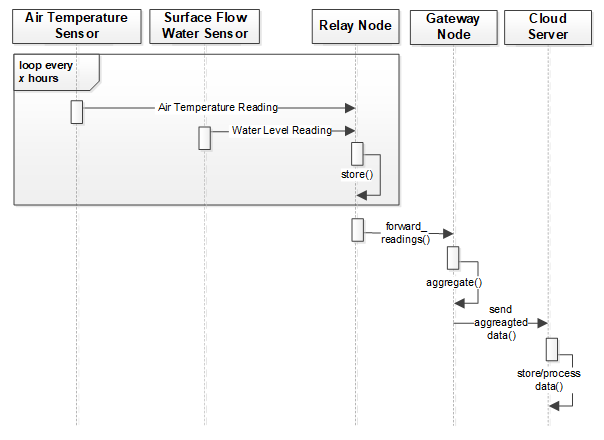}
\caption{Sequence diagram of the Environmental IoT case.}
\label{fig_EnvIoT_validationScenario}
\end{figure}

To validate the simulation environment and ensure the correctness, we traced the data from sensors to the cloud server and their data packets. We also traced the battery level of different nodes. The transfer of data between the different nodes of the architecture is illustrated in \fig{fig_EnvIoT_validationScenario}. Samples of air temperature and rainfall data traced at different nodes are listed in Table \ref{tbl_EnvIoT_validationData}. 
Such data are submitted by the sensors to the relay node and then forwarded to the gateway node. The gateway node aggregates the data of different sensors and sends it to the cloud server, which in turn stores it and produces a daily average. These are the expected behaviours for the IoT nodes considering the testbed configuration. Hence, the results reflected that the components are correctly implemented. 

\begin{table}[tb]
\caption{Sample of air temperature and rainfall data traced during the validation experiments}
\footnotesize
\label{tbl_EnvIoT_validationData}
\center
\scriptsize
\setlength{\tabcolsep}{0.7em}
\def\arraystretch{0.95}
\begin{tabular}
{
l
l
r r r
r r r
l
l
r r
r r
}
\toprule    
\multicolumn{2}{c}{\textbf{Readings}}                     &     
\multicolumn{3}{c}{\textbf{Air Sensors (\si{\degree}C)}}         &     
\multicolumn{3}{c}{\textbf{Water Sensors (mm)}}             &     
\multicolumn{2}{c}{\textbf{Relay Node}}                 &
\multicolumn{2}{c}{\textbf{Gateway Node}}             &
\multicolumn{2}{c}{\textbf{Physical Host}}                 
\\
        &
        &
S1    &        S2    &        S3        &
S4    &        S5    &        S6        &
Temp. readings        &
Precip. readings        &
Temp.        &        Precip.        &
Temp.        &        Precip.            
\\
\midrule
\parbox[t]{2mm}{\multirow{4}{*}{\rotatebox[origin=c]{90}{Day 1}}}        &
1                &
-0.34        &        -0.12        &            -0.08            &
0                &        0.01        &            0.07            &
[-0.34, -0.12, -0.08]            &
[0, 0.01, 0.07]                        &
-0.18        &        0.03        &
                &
\\
                &
2                &
-0.19        &        0.11            &            0.47            &
0.1            &        0.04        &            0.02            &
[-0.19, 0.11, 0.47]            &
[0.1, 0.04, 0.02]            &
0.13        &        0.05        &
                &
\\
                &
3                &
0.47        &        -0.87        &            -0.69            &
0.15        &        0.2            &            0.02            &
[0.47, -0.87, -0.69]            &
[0.15, 0.2, 0.02]                    &
-0.58        &        0.12        &
                &
\\
                &
4                &
0.4            &        -2.06        &            -0.54            &
0                &        0.16        &            0.14            &
[0.4, -2.06, -0.54]            &
[0    , 0.16, 0.14]                    &
-0.73        &        0.10        &
-0.34        &        0.08
\\
\midrule
\parbox[t]{2mm}{\multirow{4}{*}{\rotatebox[origin=c]{90}{Day 2}}}        &
1                &
0.28        &        0.51        &            0.49            &
0                &        0.01        &            0.07            &
[0.28, 0.51, 0.49    ]        &
[0    , 0.01, 0.07]            &
0.43        &        0.03        &
                &
\\
                &
2                &
0.33        &        0.51        &            0.74            &
0.1            &        0.04        &            0.02            &
[0.33, 0.51, 0.74]            &
[0.1, 0.04, 0.02]                &
0.53        &        0.05        &
                &
\\
                &
3                &
-0.09        &        -0.04        &            0.27            &
0.15        &        0.2            &            0.02            &
[-0.09, -0.04    , 0.27]            &
[0.15, 0.2, 0.02]                    &
0.05        &        0.12        &
                &
\\
                &
4                &
0.68        &        -0.93        &            0.09            &
0                &        0.17            &            0.14            &
[0.68, -0.93, 0.09]            &
[0, 0.17, 0.14]                    &
-0.05        &        0.10        &
0.24        &     0.08
\\
\midrule
\parbox[t]{2mm}{\multirow{4}{*}{\rotatebox[origin=c]{90}{Day 3}}}        &
1                &
1.36        &        1.71            &            1.79                &
0                &        0                &            0                    &
[1.36, 1.71, 1.79]                &
[0, 0, 0]                            &
1.62        &        0.00        &
                &
\\
                &
2                &
1.59        &        1.73            &            1.85            &
0                &        0                &            0                    &
[1.59, 1.73, 1.85]            &
[0, 0, 0]                            &
1.72            &        0.00        &
                &
\\
                &
3                &
1.11            &        1.2            &            1.5                &
0                &        0                &            0                    &
[1.11, 1.2, 1.5]                &
[0, 0, 0]                            &
1.27            &        0.00        &
                &
\\
                &
4                &
1.84        &        0.01        &            1.36            &
0                &        0                &            0                    &
[1.84, 0.01, 1.36]            &
[0, 0, 0]                    &
1.07            &        0.00        &
1.42        &        0.00    
\\
\bottomrule
\end{tabular}
\end{table}

\begin{figure*}[!th]
\centering
\subfloat[Actual Simulation Time]
    {\includegraphics[width=0.85\textwidth]{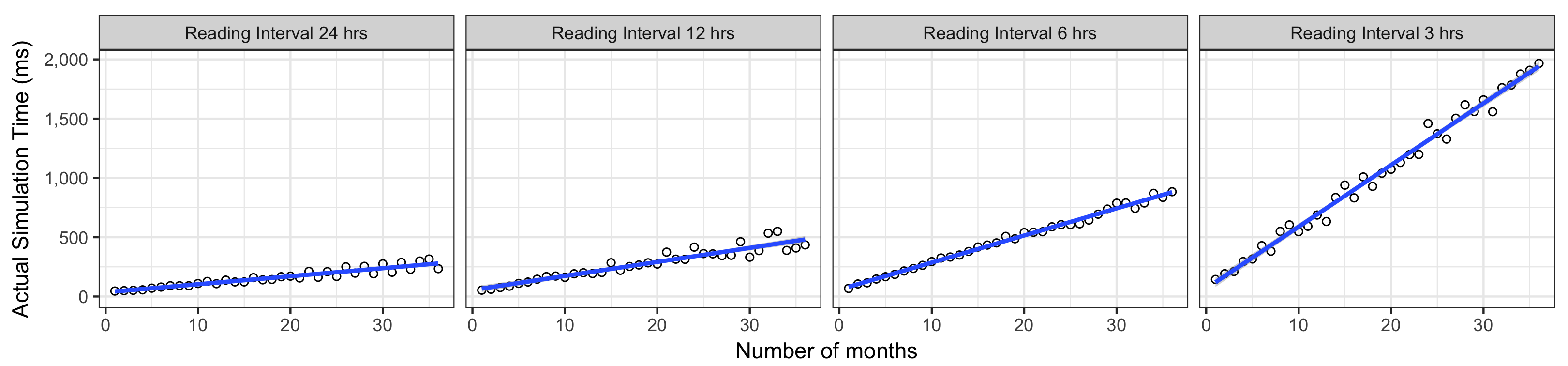}}
    \label{graph_stressEval_simTime}
\subfloat[Used Memory]
    {\includegraphics[width=0.85\textwidth]{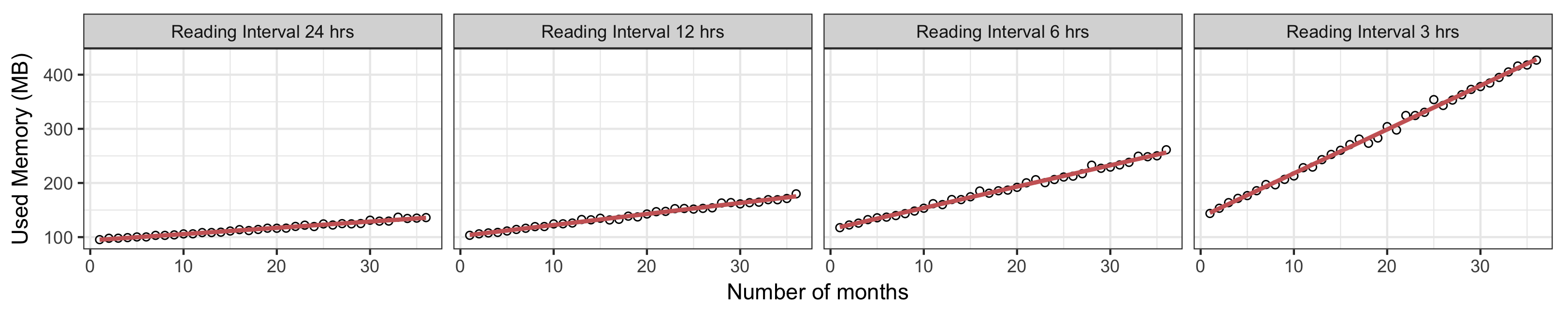}}
    \label{graph_stressEval_usedMemory}
\caption{Simulation overhead measured as the time of the simulation increases, each point represents the mean of 30 runs.}
\label{graph_stressEval}
\end{figure*}

\paragraph*{\textbf{Evaluation experiments and results}}
To evaluate the capability of the simulation platform, we ran two sets of experiments where we vary the number of days and the number of nodes simulated. These two variables represent the scale of an IoT system and its runtime. 
Both sets are based on different reading interval from the sensors. The variation in the reading interval reflects different possible scenarios. For instance, a heavy rainy season with expected flooding would require more frequent sensor readings, while dry seasons require less frequent readings. We used four different reading intervals in our experiments, which are 24, 12, 6 and 3 hours. The reading intervals are indirectly proportional to the number of events in the simulation process to stress test the platform. 

In the first set of experiments, we varied the number of days simulated. We have setup one testbed specified in Table \ref{tbl_EnvIoT_testbedConfiguration} for the different reading intervals. We have used readings of the most recent 3 years of the dataset (from the beginning of January 2014 to the end of December 2016). We measured the actual simulation time and memory footprint for a varying simulation period (1 to 36 months). 

\fig{graph_stressEval} shows the actual simulation time and memory footprint respectively for the stress experiments. The results show the trend of both factors increasing by the simulation period. The trend is also varying clearly according to the reading intervals. The variations in the simulation period and reading interval helped to stress test the environment, without requiring a high-performance computer. Additionally, both behaviours are expected which reflects the correctness of our implementation.

In the second set of experiments, we scaled the testbed by the number of locations, where each location contains a full setup of the testbed specified in Table~\ref{tbl_EnvIoT_testbedConfiguration}. 
We varied the number of locations between 1 and 100, \ie up to 900 nodes. We examined the associated simulation overhead for different reading intervals for a simulation period of one month. 

\begin{figure*}[!h]
\centering
\subfloat[Actual Simulation Time]
    {\includegraphics[width=0.90\textwidth]{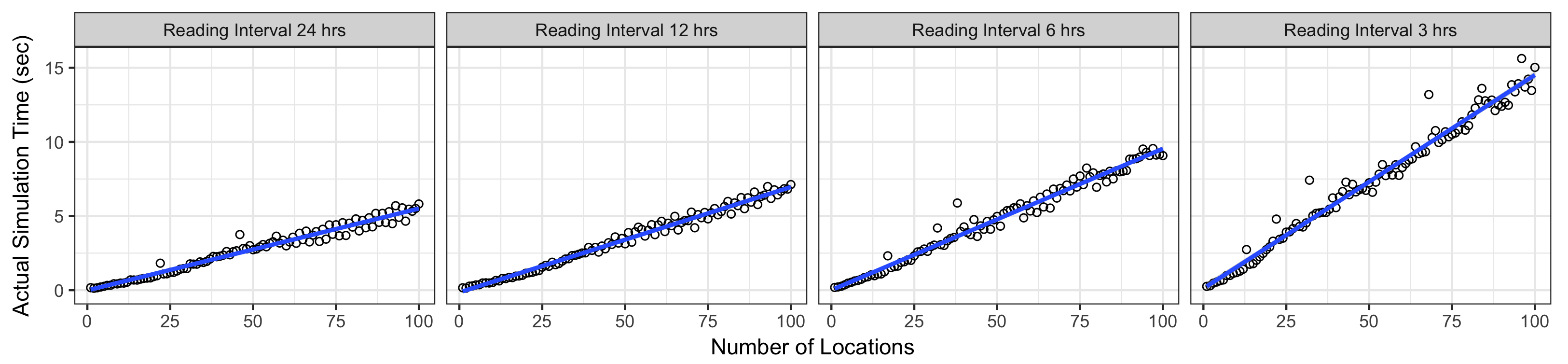}}
    \label{graph_scalabilityEval_simTime}
\subfloat[Used Memory]
    {\includegraphics[width=0.90\textwidth]{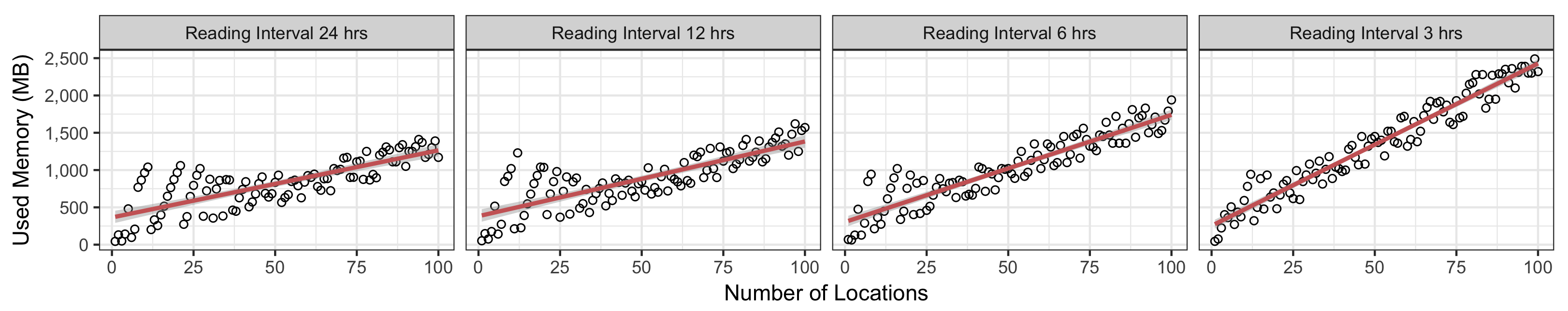}}
    \label{graph_scalabilityEval_usedMemory}
\caption{Simulation overhead measured as the number of locations increases, each point represents the mean of 30 runs.}
\label{graph_scalabilityEval}
\end{figure*}

\fig{graph_scalabilityEval} illustrates the actual simulation time and memory footprint. As expected, the simulation time proportionally increases with the number of locations and reading intervals. Meanwhile, the memory footprint shows more fluctuations around the regression line, which could be attributed to the large variety of data transferred between a large number of nodes.

\subsection{Case Study: An End-to-End IoT service using JOSE testbed}
\label{sec_evaluation_JOSE}
\begin{figure}[!h]
\centering
\includegraphics[width=0.99\columnwidth]{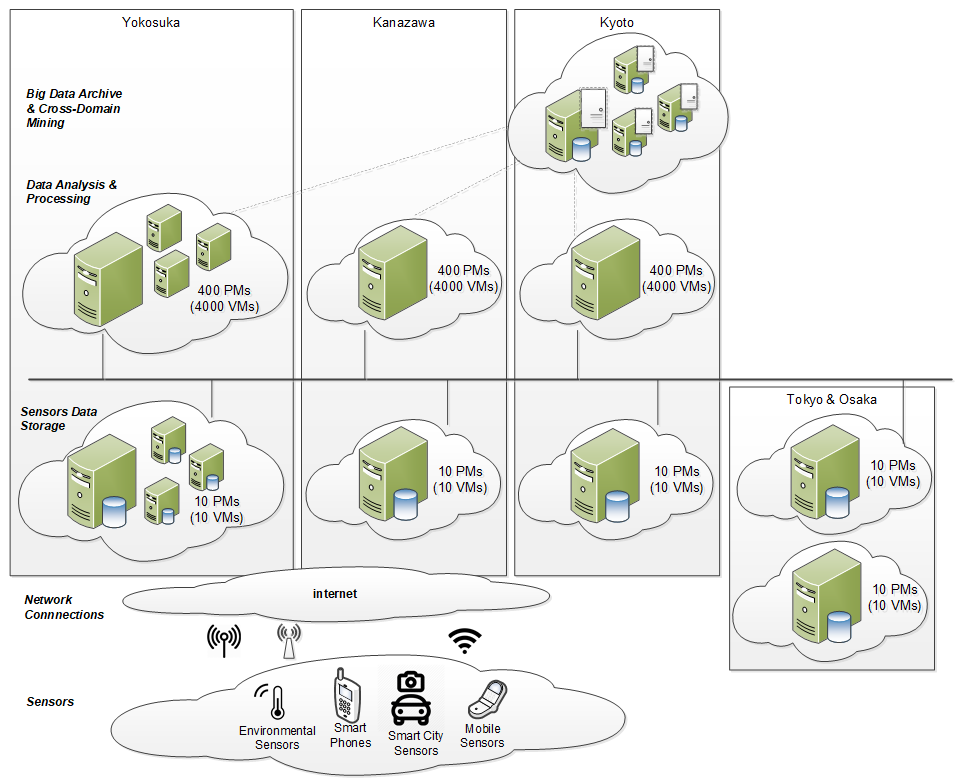}
\caption{An architectural view of the JOSE testbed (simplified and redrawn from \cite{Teranishi2015}).}
\label{fig_JoseTestbed}
\end{figure}

The Japan-wide Orchestrated Smart/Sensor Environment (JOSE) is an open testbed and service platform that can accommodate large-scale IoT services \cite{Teranishi2015}. It was developed as an IaaS for IoT services, distributed in 5 cities around Japan. The architecture of the testbed is illustrated in \fig{fig_JoseTestbed}. The infrastructure consists of: (i) a huge number of environmental, mobile and camera sensors; (ii) multiple types of wireless sensor networks; (iii) gateways for each field trials; (iv) data storage resources at each location; and (v) computational resources for analysing and processing data for each sensor type. The storage resources are distributed at all 5 locations, where each location has 10 PMs, while the computational resources are distributed at 3 locations, where each location has 400 servers. For both types of resources, each PM can run 10 VMs.

We simulated the whole testbed (\ie 5 locations), a total of 50 PMs with 500 VMs for storage, and 1200 PMs with 12000 VMs for computation. Since the configurations of PMs and VMs in JOSE are not available, we configured PMs in our simulation as IBM x3550 servers each of 2x Xeon X5670 3GHz, 6 cores and 256GB RAM. VM characteristics correspond to the latest generation of General Purpose Amazon EC2 instances~\cite{AmazonEC2}, from which we used \texttt{m4.large} (2 core vCPU 2.4GHz, 8GB RAM). In this case study, we included more sensor types. In particular, we used air temperature, air humidity, air pressure, wind speed, and precipitation sensors. For the sensors data, we used historical datasets from the UK Centre for Environmental Data Analysis \cite{MetOffice2017}.

We configured an end-to-end IoT service for disaster monitoring which is responsible for sending alerts of varying types given certain thresholds for each measurement (inspired by \cite{Hernandez2016}). For instance, a flood monitoring service would give a \textit{green} alert when precipitation readings are rising, \textit{yellow} when they continue to rise, and \textit{red} when they reach the flooding threshold. Using this infrastructure, we conducted several experiments by varying the number of sensors of each type at each city. For instance, we configured the first set of experiments with 1,000 sensors for each of the 5 sensor types at each city, \ie a total of 25,000 sensors. We conducted each experiment at different reading intervals as in the environment IoT case study, varying the experiment duration (from 1 to 12 months) and the number of sensors (from 1,000 to 1,000,000). 

\begin{figure*}[!th]
\centering
\subfloat[Actual Simulation Time]
    {\includegraphics[width=0.90\textwidth]{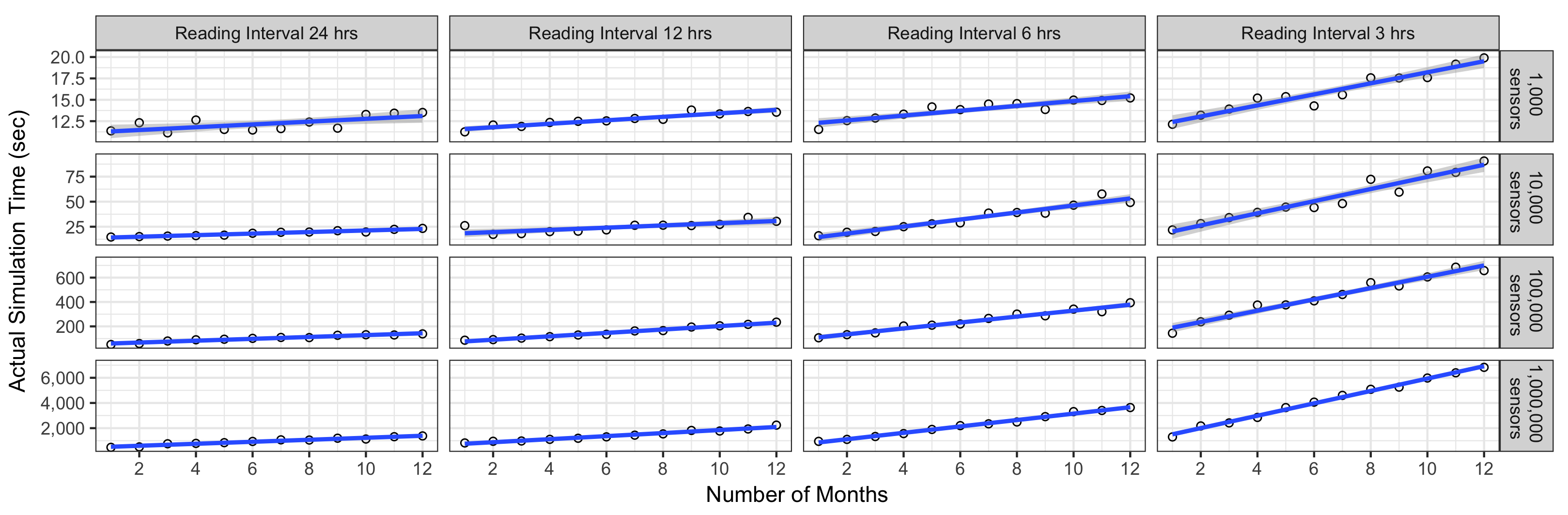}}
    \label{graph_JoseEval_simTime}
\subfloat[Used Memory]
    {\includegraphics[width=0.90\textwidth]{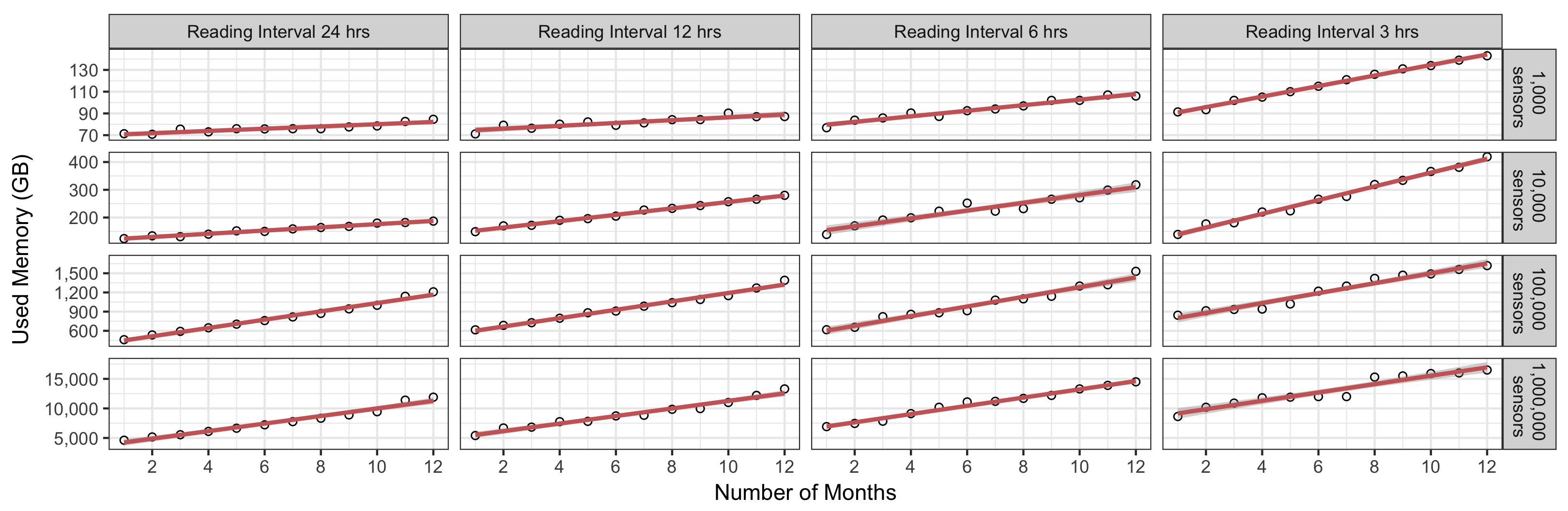}}
    \label{graph_JoseEval_usedMemory}
\caption{Results of simulating the JOSE testbed, each point represents the mean of 30 runs.}
\label{graph_JoseEval}
\end{figure*}

\fig{graph_JoseEval} depicts the actual simulation time and used memory for different experiments. All three factors (\ie simulation duration, scale, and reading frequency) increase simulation overhead with a strong linear trend. Comparing the results of the two case studies, the simulation overhead varies in the same trend when using the same reading interval. 


\section{Discussion}
\label{sec_discussion}
The proposed IoTNetSim platform supports modelling and simulation of end-to-end IoT systems, including the different layers that an IoT system might be composed of (\ie cloud, fog, edge). The platform features fine-grained modelling of IoT nodes and associated networking with different types of connections and area coverage, as well as a multitude of application and device models.

We applied IoTNetSim in two real cases of contrasting features. The first case study is of an IoT deployment for the natural environment, a current common scenario of IoT services. Starting with the original small-scale deployment, we conducted a number of experiments where we varied the experimentation period, scale and settings. IoTNetSim exhibited very modest overhead in all configurations. Our second case study features JOSE, where IoTNetSim was able to fully model and simulate such extremely large-scale IoT infrastructure. The overheads were considerably higher than in the first use case, due to the nature of the simulated testbed. We argue that the environment presents a reasonable benefit-cost ratio in terms of enabling a thorough evaluation of such large-scale system prior to deployment (which would cost the vicinity of \$3m as a conservative estimate).

Based on the modelling and simulation challenges in IoT by Kecskemeti \etal \cite{Kecskemeti2017}, we argue that our work tackled the following: (i) the ability of modelling device heterogeneity, that is accommodating diverse sensor and actuator types with their fine-grained details, as well as resulting data types and sizes; (ii) modelling different population of sensors and actuators, where researchers can conduct what-if scenarios to study the placement, population and density of sensing devices and their effect on the accuracy of data analysed; (iii) realistic parametrisation of IoT simulation models, such as reading intervals, network signals, devices failure, battery consumption; (iv) support for online decision making, where researchers can incorporate live data and process multiple scenarios. Also, the platform is customisable and extendable, supporting new device models and different experiments. 

Meanwhile, there are potential threats to validity of the proposed work. First, the choice of the evaluation case studies and the testbed configurations could be biased by the authors' background and knowledge. Our mitigation strategy for this issue was to base the case studies on previous work \cite{Nundloll2016} and actual testbeds \cite{Teranishi2015} by a different group of researchers, which makes us believe that the evaluation setup is practical and challenging. Second, the fact that the proposed work is evaluated by its authors presents a threat to objectivity. To mitigate this risk, we sought to conduct other sets of experiments with other researchers in order to ensure wide feasibility of the platform. Third, there is a lack of validating the simulation accuracy against the real system in a systematic manner. Yet, the results reflect the expected behaviour of the simulation.

While the simulation environment supports scenarios of network failures and/ or signal loss, we have not included such scenarios in our initial evaluation. The simulation environment is also able to support a wide range of sensors, but we have not included all such types (\eg mobile sensors) in our experiments due to the complexity of moving their locations. Such scenarios require evaluation using multiple hypothetical scenarios, which is beyond our scope here.

\section{Related Work}
\label{sec_relatedWork}
Table \ref{tbl_relatedWorkComparison} exhibits a comparative summary of the features of the simulation environments discussed below.

\begin{table*}[!ht]
\def\arraystretch{0.9}
\caption{Current simulators used for IoT research sorted chronologically and compared against our IoTNetSim.}%
\label{tbl_relatedWorkComparison}
\center
\footnotesize
\begin{tabular}
{lccccccc}
\toprule
& &     
{\textbf{Application}}                    &
&
{\textbf{Fog/Edge}}                        &
&
{\textbf{Networking-}}             &
{\textbf{Mobility-}}                     
\\
{\textbf{Simulator}}                                 &     
{\textbf{Ref.}}                                 &     
{\textbf{Layer}}                    &
{\textbf{Cloud Layer}}                            &
{\textbf{Layer}}                        &
{\textbf{IoT Nodes}}                                 &
{\textbf{support}}             &
{\textbf{support}}                     
\\
\midrule
PlanetSim         &        \cite{Garcia2005}            &
\xmark            &
\cmark            &
\xmark            &
\xmark            &
\cmark            &
\xmark            
\\
\rowcolor{Gray} 
J-Sim             &        \cite{Sobeih2006}                        &
\xmark            &
\xmark            &
\xmark            &
\xmark            &
\cmark            &
\cmark            
\\
OverSim         &        \cite{Baumgart2007}            &
\xmark            &
\cmark            &
\xmark            &
\xmark            &
\cmark            &
\xmark            
\\
\rowcolor{Gray} 
OMNeT++     &        \cite{Varga2008}                &
\xmark            &
\xmark            &
\xmark            &
\xmark            &
\cmark            &
\cmark            
\\
MDCSim         &        \cite{Lim2009}                        &
\cmark            &
\cmark            &
\xmark            &
\xmark            &
\cmark            &
\xmark            
\\
\rowcolor{Gray} 
NS-3                 &        \cite{Riley2010}                            &
\xmark            &
\xmark            &
\xmark            &
\xmark            &
\cmark            &
\cmark            
\\
CloudSim         &        \cite{Calheiros2011}            &
\cmark            &
\cmark            &
\xmark            &
\xmark            &
\xmark            &
\xmark            
\\
\rowcolor{Gray} 
GreenCloud &        \cite{Kliazovich2012}            &
\cmark            &
\cmark            &
\xmark            &
\xmark            &
\cmark            &
\xmark            
\\
iCanCloud     &        \cite{Nunez2012}                &
\cmark            &
\cmark            &
\xmark            &
\xmark            &
\cmark            &
\xmark            
\\
\rowcolor{Gray} 
SimIoT             &            \cite{Sotiriadis2014}                 & 
\cmark            &
\cmark            &
\xmark            &
\xmark            &
\xmark            &
\xmark            
\\
iFogSim         &        \cite{Gupta2016}        & 
\cmark            &
\cmark            &
\cmark            &
\xmark            &
\xmark            &
\xmark            
\\
\rowcolor{Gray}
IOTSim             &        \cite{Zeng2017}                    & 
\cmark            &
\cmark            &
\xmark            &
\cmark            &
\xmark            &
\xmark            
\\
\textbf{IoTNetSim}            &    [this]        &
\cmark            &
\cmark            &
\cmark            &
\cmark            &
\cmark            &
\cmark            
\\
\bottomrule
\end{tabular}
\end{table*}

A recent survey \cite{Chernyshev2018} categorised \textbf{IoT simulators} into ones that support \textit{IoT elements} and those that focus on \textit{IoT applications}. An example of the former category is iFogSim that focuses on resource management for fog and edge computing \cite{Gupta2016} but lacks modelling IoT nodes and support for networking. Examples of the latter category include IOTSim for modelling and analysing IoT applications and big data processing \cite{Zeng2017} and SimIoT for evaluating job processing times in cloud-based systems based on sensor data \cite{Sotiriadis2014}. The survey also considered general-purpose network simulators that evolved to support IoT-specific components. 

\textbf{Network simulators} such as J-Sim~\cite{Sobeih2006}, OMNeT++~\cite{Varga2008} and NS-3~\cite{Riley2010} have been extensively used in wireless sensor networks research. Peer-to-peer and overlay networks have been simulated in OverSim \cite{Baumgart2007} and PlanetSim \cite{Garcia2005} respectively. However, the lack of built-in support for IoT standards, IoT-specific radio models and application-level protocols limits their practical applicability for IoT research.

Notable \textbf{cloud simulators} include: 
CloudSim~\cite{Calheiros2011}, a modular simulator of large scale clouds; 
GreenCloud~\cite{Kliazovich2012}, a packet-level simulator of energy-aware cloud data centers; 
MDCSim~\cite{Lim2009} for simulating multi-tier data centres in detail; and iCanCloud~\cite{Nunez2012} which focuses on modelling flexibility and scalability. 
Other tools specialise in simulating specific issues, such as power consumption and scientific workflows processing and containers in cloud data centers \cite{Sakellari2013}. 
Some cloud simulators have been extended for modelling fog and edge computing, yet they are limited in supporting IoT nodes.


\section{Conclusion and Future Work}
\label{sec_conclusion}
We presented IoTNetSim, a novel platform for end-to-end modelling and simulation of IoT systems and services, \ie from data sensing phase to data analysis in the cloud. The self-contained environment embeds modules for detailed modelling of IoT nodes, supports different designs of IoT networks, and simulates scenarios of network and battery failures. We demonstrate through experimental evaluation that IoTNetSim is able to model and simulate large-scale IoT infrastructures and different IoT services with reasonable overhead.

To complement the evaluation, we plan to evaluate the accuracy of the simulation results. We also plan to conduct further evaluations of IoT services from different domains, such as smart home and digital health, as well as case studies that feature mobile sensors (\eg smartphones and animal trackers). Our future work will focus on automating the placement of IoT nodes using different algorithms. 
Such extension helps in designing IoT architectures for more efficient data collection and topology design.

 \section*{Acknowledgments}

This work was funded by EPSRC Grant EP/M015734/1: `Declarative and Interoperable Overlay Networks, Applications to Systems of Systems (\textit{DIONASYS})'. 
The work was also partially supported by EPSRC Grant EP/N027736/1: `\textit{Models in the Cloud}: Generative Software Frameworks to Support the Execution of Environmental Models in the Cloud'.

\bibliographystyle{ieeetr}
\bibliography{IoTsim-full}

\begin{thebibliography}{10}

\bibitem{Chernyshev2018}
M.~Chernyshev, Z.~Baig, O.~Bello, and S.~Zeadally, ``Internet of things
  ({IoT}): Research, simulators, and testbeds,'' {\em IEEE Internet of Things
  Journal}, vol.~5, no.~3, pp.~1637--1647, 2018.

\bibitem{Buyya2009}
R.~Buyya, R.~Ranjan, and R.~N. Calheiros, ``Modeling and simulation of scalable
  cloud computing environments and the cloudsim toolkit: Challenges and
  opportunities,'' in {\em Conference on High Performance Computing \&
  Simulation (HPCS)}, pp.~1--11, IEEE, 2009.

\bibitem{Armbrust2010}
M.~Armbrust, A.~Fox, R.~Griffith, A.~D. Joseph, R.~Katz, A.~Konwinski, G.~Lee,
  D.~Patterson, A.~Rabkin, I.~Stoica, and M.~Zaharia, ``A view of cloud
  computing,'' {\em ACM Communications}, vol.~53, no.~4, pp.~50--58, 2010.

\bibitem{Calheiros2011}
R.~N. Calheiros, R.~Ranjan, A.~Beloglazov, C.~De~Rose, and R.~Buyya,
  ``{CloudSim}: a toolkit for modeling and simulation of cloud computing
  environments and evaluation of resource provisioning algorithms,'' {\em
  Software: Practice and Experience}, vol.~41, no.~1, pp.~23--50, 2011.

\bibitem{Quiroz2009}
A.~Quiroz, K.~Hyunjoo, M.~Parashar, N.~Gnanasambandam, and N.~Sharma, ``Towards
  autonomic workload provisioning for enterprise grids and clouds,'' in {\em
  IEEE/ACM Conference on Grid Computing}, pp.~50--57, 2009.

\bibitem{Misra2017}
S.~Misra, M.~Maheswaran, and S.~Hashmi, {\em System Model for the Internet of
  Things}, pp.~5--17.
\newblock Springer International Publishing, 2017.

\bibitem{AlFuqaha2015}
A.~Al-Fuqaha, M.~Guizani, M.~Mohammadi, M.~Aledhari, and M.~Ayyash, ``Internet
  of things: A survey on enabling technologies, protocols, and applications,''
  {\em IEEE Communications Surveys \& Tutorials}, vol.~17, no.~4,
  pp.~2347--2376, 2015.

\bibitem{Vermesan2011}
O.~Vermesan, P.~Friess, P.~Guillemin, S.~Gusmeroli, H.~Sundmaeker, A.~Bassi,
  I.~S. Jubert, M.~Mazura, M.~Harrison, and M.~Eisenhauer, ``Internet of things
  strategic research roadmap,'' {\em Internet of Things-Global Technological
  and Societal Trends}, vol.~1, pp.~9--52, 2011.

\bibitem{Papadopoulos2017}
G.~Z. Papadopoulos, A.~Gallais, G.~Schreiner, E.~Jou, and T.~Noel, ``Thorough
  {IoT} testbed characterization: From proof-of-concept to repeatable
  experimentations,'' {\em Computer Networks}, vol.~119, pp.~86--101, 2017.

\bibitem{Papadopoulos2013}
G.~Z. Papadopoulos, J.~Beaudaux, A.~Gallais, T.~Noël, and G.~Schreiner,
  ``Adding value to wsn simulation using the iot-lab experimental platform,''
  in {\em IEEE Conference on Wireless and Mobile Computing, Networking and
  Communications (WiMob)}, pp.~485--490, 2013.

\bibitem{Kecskemeti2017}
G.~Kecskemeti, G.~Casale, D.~N. Jha, J.~Lyon, and R.~Ranjan, ``Modelling and
  simulation challenges in internet of things,'' {\em IEEE Cloud Computing},
  vol.~4, no.~1, pp.~62--69, 2017.

\bibitem{Kathiravelu2014a}
P.~Kathiravelu and L.~Veiga, ``Concurrent and distributed {CloudSim}
  simulations,'' in {\em IEEE Symposium on Modelling, Analysis \& Simulation of
  Computer and Telecommunication Systems (MASCOTS)}, pp.~490--493, 2014.

\bibitem{Garg2011}
S.~K. Garg and R.~Buyya, ``{NetworkCloudSim}: Modelling parallel applications
  in cloud simulations,'' in {\em IEEE Conference on Utility and Cloud
  Computing (UCC)}, pp.~105--113, 2011.

\bibitem{elkhatib2017microclouds}
Y.~Elkhatib, B.~F. Porter, H.~B. Ribeiro, M.~F. Zhani, J.~Qadir, and
  E.~Rivière, ``On using micro-clouds to deliver the fog,'' {\em Internet
  Computing}, vol.~21, no.~2, pp.~8--15, 2017.

\bibitem{Nundloll2016}
V.~Nundloll, B.~Porter, G.~S. Blair, B.~Emmett, J.~Cosby, D.~L. Jones,
  D.~Chadwick, B.~Winterbourn, P.~Beattie, G.~Dean, R.~Shaw, W.~Shelley,
  M.~Brown, and I.~Ullah, ``The design and deployment of an end-to-end {IoT}
  infrastructure for the natural environment,'' {\em Future Internet}, vol.~11,
  no.~6, 2019.

\bibitem{MetOffice2017}
D.~Hollis and M.~McCarthy, ``{UKCP09: Met Office} gridded and regional land
  surface climate observation datasets.''
  \url{http://catalogue.ceda.ac.uk/uuid/87f43af9d02e42f483351d79b3d6162a},
  2017.
\newblock Centre for Environmental Data Analysis.

\bibitem{Teranishi2015}
Y.~Teranishi, Y.~Saito, S.~Murono, and N.~Nishinaga, ``{JOSE}: An open testbed
  for field trials of large-scale {IoT} services,'' {\em Journal of the
  National Institute of Information and Communications Technology}, vol.~62,
  no.~2, 2015.

\bibitem{AmazonEC2}
{\relax Amazon Web Services, Inc.}, ``{\relax Amazon EC2 Instance Types}.''
\newblock Accessed: 2017-10-01.

\bibitem{Hernandez2016}
J.~A. Hernández-Nolasco, M.~A.~W. Ovando, F.~D. Acosta, and P.~Pancardo,
  ``Water level meter for alerting population about floods,'' in {\em IEEE
  Conference on Advanced Information Networking and Applications (AINA)},
  pp.~879--884, 2016.

\bibitem{Garcia2005}
P.~Garc\'ia, C.~Pairot, R.~Mondéjar, J.~Pujol, H.~Tejedor, and R.~Rallo,
  ``{PlanetSim}: A new overlay network simulation framework,'' in {\em Software
  Engineering and Middleware} (T.~Gschwind and C.~Mascolo, eds.), pp.~123--136,
  Springer Berlin Heidelberg, 2005.

\bibitem{Sobeih2006}
A.~Sobeih, J.~C. Hou, K.~Lu-Chuan, L.~Ning, Z.~Honghai, C.~Wei-Peng,
  T.~Hung-Ying, and L.~Hyuk, ``{J-Sim}: A simulation and emulation environment
  for wireless sensor networks,'' {\em IEEE Wireless Communications}, vol.~13,
  no.~4, pp.~104--119, 2006.

\bibitem{Baumgart2007}
I.~Baumgart, B.~Heep, and S.~Krause, ``{OverSim}: A flexible overlay network
  simulation framework,'' in {\em IEEE Global Internet Symposium}, pp.~79--84,
  2007.

\bibitem{Varga2008}
A.~Varga and R.~Hornig, ``An overview of the {OMNeT++} simulation
  environment,'' in {\em Conference on Simulation Tools and Techniques for
  Communications, Networks and Systems}, pp.~1--10, ICST, 2008.

\bibitem{Lim2009}
S.~H. Lim, B.~Sharma, G.~Nam, E.~K. Kim, and C.~R. Das, ``{MDCSim}: A
  multi-tier data center simulation, platform,'' in {\em IEEE Conference on
  Cluster Computing and Workshops}, pp.~1--9, 2009.

\bibitem{Riley2010}
G.~F. Riley and T.~R. Henderson, {\em The ns-3 Network Simulator}, pp.~15--34.
\newblock Berlin, Heidelberg: Springer Berlin Heidelberg, 2010.

\bibitem{Kliazovich2012}
D.~Kliazovich, P.~Bouvry, and S.~U. Khan, ``{GreenCloud}: A packet-level
  simulator of energy-aware cloud computing data centers,'' {\em The Journal of
  Supercomputing}, vol.~62, no.~3, pp.~1263--1283, 2012.

\bibitem{Nunez2012}
A.~Nunez, J.~L. Vazquez-Poletti, A.~C. Caminero, G.~G. Castane, J.~Carretero,
  and I.~M. Llorente, ``{iCanCloud}: A flexible and scalable cloud
  infrastructure simulator,'' {\em Journal of Grid Computing}, vol.~10, no.~1,
  pp.~185--209, 2012.

\bibitem{Sotiriadis2014}
S.~Sotiriadis, N.~Bessis, E.~Asimakopoulou, and N.~Mustafee, ``Towards
  simulating the internet of things,'' in {\em Conference on Advanced
  Information Networking and Applications Workshops}, pp.~444--448, 2014.

\bibitem{Gupta2016}
H.~Gupta, A.~V. Dastjerdi, S.~K. Ghosh, and R.~Buyya, ``{iFogSim}: A toolkit
  for modeling and simulation of resource management techniques in internet of
  things, edge and fog computing environments,'' {\em Software: Practice and
  Experience}, vol.~47, no.~9, pp.~1275--1296, 2017.

\bibitem{Zeng2017}
X.~Zeng, S.~K. Garg, P.~Strazdins, P.~P. Jayaraman, D.~Georgakopoulos, and
  R.~Ranjan, ``{IOTSim}: A simulator for analysing iot applications,'' {\em
  Journal of Systems Architecture}, vol.~72, pp.~93--107, 2017.

\bibitem{Sakellari2013}
G.~Sakellari and G.~Loukas, ``A survey of mathematical models, simulation
  approaches and testbeds used for research in cloud computing,'' {\em
  Simulation Modelling Practice and Theory}, vol.~39, no.~0, pp.~92--103, 2013.

\end{thebibliography}

\end{document}